\documentclass{elsart} 

\usepackage{amsmath}
\usepackage{epsfig}
\usepackage{amssymb}

\journal{Physics Letters B }

\newcommand{\brproduct}{${\cal B}(\bar{B}^0\to\Sigma_c(2455)^{0}\bar{N}^0)\times{\cal B}(\bar{N}^0\to\bar{p}\pi^+)$=({\brscn})$\times10^{-4}$}
\newcommand{\nbrlcpiapp}{$2.1\pm0.2\rm{(stat.)}\pm0.3\rm{(syst.)}\pm0.5$} 
\newcommand{\nbrlcpiazr}{$1.4\pm0.2\rm{(stat.)}\pm0.2\rm{(syst.)}\pm0.4$} 

\newcommand{\brscn}{$0.80\pm0.15\rm{(stat.)}\pm0.14\rm{(syst.)}\pm0.21$}
\newcommand{\yieldp}{$70\pm11$}
\newcommand{\yieldl}{$32\pm9$}
\newcommand{\sgnfn}{7.0}  
\newcommand{\ssgnfn}{6.1} 
\newcommand{\sgnfl}{4.6}

\newcommand{\yieldss}{$71\pm11\pm10$}
\newcommand{\yieldsp}{$70\pm11\pm10$}

\newcommand{\masss}{$1473\pm31\pm2$}
\newcommand{\widths}{$315\pm72\pm53$}
\newcommand{\massp}{$1516\pm29\pm14$}
\newcommand{\widthp}{$365\pm97\pm90$}

\newcommand{\thbody}{$\bar{B}^0\to\Sigma_{c}^{0}\bar{p}\pi^+$}
\newcommand{\MCA}{$\bar{B}^0\to\Sigma_{c}^{0}\bar{N}^0$}
\newcommand{\MCB}{$\bar{B}^0\to\Sigma_{c}^{0}\bar{p}\pi^+$}
\newcommand{\MCC}{$\bar{B}^0\to (\Sigma^0_c\pi^+)_{\rm X}\,\bar{p}$}
\newcommand{\lcx}{$(\Sigma_c^0\pi^+)_{\rm X}$}

\newcommand{\lcknown}{$\Lambda_c^{*+}$}

\usepackage{graphicx} 
\usepackage{dcolumn}  

\begin{document}

\resizebox{!}{3cm}{\includegraphics{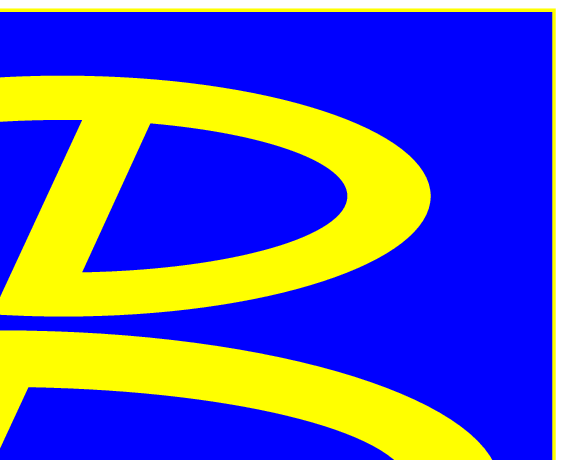}}
\hspace{5cm}{  \vbox{
               \hbox{ }
               \hbox{Belle Preprint 2008-23}
               \hbox{KEK Preprint   2008-23}
               \hbox{ }
               \hbox{ }
               \hbox{ }
}}

\begin{frontmatter}

\title{  
Study of intermediate two-body decays 
in $\bar{B}^0\to \Sigma_c(2455)^{0}\bar{p}\pi^{+}$
}
\collab{Belle Collaboration}
   \author[Kyungpook]{H.~O.~Kim\corauthref{cor}}, 
   \ead{hokim@knu.ac.kr}
   \author[KEK]{H.~Kichimi\corauthref{cor}}, 
   \ead{kichimi@post.kek.jp}
   \corauth[cor]{Corresponding author.}
   \author[KEK]{I.~Adachi}, 
   \author[Tokyo]{H.~Aihara}, 
  \author[BINP]{K.~Arinstein}, 
  \author[BINP]{V.~Aulchenko}, 
   \author[Lausanne,ITEP]{T.~Aushev}, 
   \author[Sydney]{A.~M.~Bakich}, 
   \author[ITEP]{V.~Balagura}, 
  \author[BINP]{I.~Bedny}, 
   \author[BINP]{A.~Bondar}, 
   \author[Krakow]{A.~Bozek}, 
% ...

\address[BINP]{Budker Institute of Nuclear Physics, Novosibirsk, Russia}
% ...
\address[KEK]{High Energy Accelerator Research Organization (KEK), Tsukuba, Japan}
% ...
\address[ITEP]{Institute for Theoretical and Experimental Physics, Moscow, Russia}
% ...
\address[Kyungpook]{Kyungpook National University, Taegu, South Korea}
\address[Lausanne]{\'Ecole Polytechnique F\'ed\'erale de Lausanne, EPFL, Lausanne, Switzerland}
% ...
\address[Krakow]{H. Niewodniczanski Institute of Nuclear Physics, Krakow, Poland}
% ...
\address[Sydney]{University of Sydney, Sydney, NSW, Australia}
% ...
\address[Tokyo]{Department of Physics, University of Tokyo, Tokyo, Japan}
% ...


\begin{abstract}
We present results of a detailed study of the three-body 
$\bar{B}^0\to\Sigma_c(2455)^{0}\bar{p}\pi^{+}$ decay. A significant enhancement of
signal events is observed in the $\bar{p}\pi^+$ mass system near
$1.5\,{\rm GeV}/c^2$ that is consistent with the presence of 
an intermediate baryonic resonance $\bar{N}^0$, where $\bar{N}^0$ is the $\bar{N}(1440)^0 P_{11}$
or $\bar{N}(1535)^0 S_{11}$ state, or an admixture of the two states. We measure
the product {\brproduct},
where the last error is due to the uncertainty in ${\cal B}(\Lambda_c^+\to p K^-\pi^+)$.
The significance of the signal is {\ssgnfn} standard deviations.
This analysis is based on a data sample of 357 fb$^{-1}$, accumulated at the
$\Upsilon(4S)$ resonance with the Belle detector at the KEKB asymmetric-energy $e^+e^-$
collider.  
\end{abstract}

\end{frontmatter}




\normalsize

\newpage

\normalsize

\vskip 1.0cm

Various charmed baryonic $B$ decays into four-, three- and two-body final states have been reported
~\cite{park,gaby1,gaby2,gaby3,gaby4,cleo_lamc,cleo_blamc}, and
the measured branching fractions show clearly that the branching  
fraction increases with the multiplicity of the final state~\cite{beach04,gaby2}. 
To understand this hierarchy, it is interesting to study decays of 
$\bar{B}^0\to\Lambda_c^+{\bar p}\pi^+\pi^-$ into three- and two-body final states.
The branching fractions are predicted from CKM matrix elements~\cite{ckm}, while the form factors 
of the decay vertices depend on the decay mechanism.
Experimental studies provide stringent constraints on the theoretical models
~\cite{jarfi,chernyak,lamc2_last_theory}.

In this report, we perform a detailed study of the intermediate three-body decay  
$\bar{B}^0\to\Sigma_c(2455)^0\bar{p}\pi^+$ observed in the previous analysis of 
$\bar{B}^0\to\Lambda_c^+\bar{p}\pi^+\pi^-$~\cite{park},
using a data sample of $388\times10^6$ $B\bar B$ events, corresponding to 357 fb$^{-1}$
accumulated at the $\Upsilon(4S)$ resonance with the Belle detector at
the KEKB asymmetric-energy $e^+e^-$ collider~\cite{kekb}.  

The Belle detector is a large-solid-angle spectrometer based on 
a 1.5~Tesla superconducting solenoid magnet. It consists of a silicon vertex detector (SVD) (a three-layer 
SVD for the first sample of $(152.0 \pm 1.2)\times10^6$ $B\bar{B}$ events
and a four-layer SVD for the latter $(235.8 \pm 3.6)\times10^6$ $B\bar{B}$ events),
a 50-layer central drift chamber (CDC), an array of aerogel threshold Cherenkov counters (ACC),
a barrel-like arrangement of time-of-flight scintillation counters (TOF), and an electromagnetic
calorimeter (ECL) comprised of CsI\,(Tl) crystals located inside the superconducting 
solenoid coil. An iron flux return located outside the coil is instrumented to detect $K_L^0$ mesons
and to identify muons (KLM). The detector is described in detail elsewhere~\cite{belle}. 
We simulate the detector response and estimate the efficiency for signal reconstruction by Monte Carlo simulation (MC). 
We use the EvtGen program~\cite{evtgen} for signal event generation and
a GEANT-based~\cite{geant} detector simulation program to model the Belle detector response for the signal.

\begin{figure}
\centering
\includegraphics[width=0.50\textwidth]{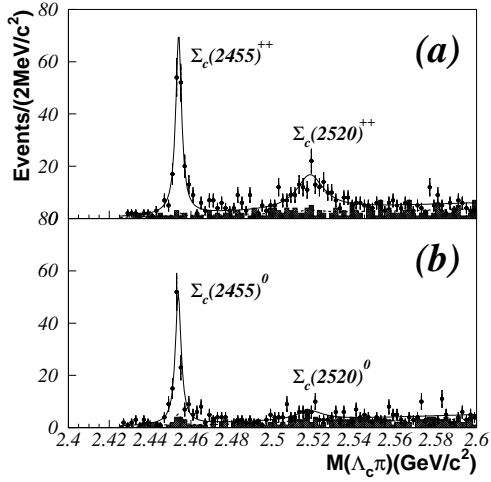}
\centering
\caption{
The mass distributions of (a) $\Lambda_{\mathrm{c}}^+\pi^+$ and 
(b) $\Lambda_{\mathrm{c}}^+\pi^-$ in  $\bar{B}^0\rightarrow\Lambda_{\mathrm{c}}^+\bar{p}\pi^+\pi^-$.
The points with error bars show the mass distribution for the events in the $B$ signal box, and
the shaded histogram indicates that for the background.
The solid and dashed curves represent the signal and the background, respectively,
obtained from a simultaneous binned likelihood fit.
}
\label{fig:4-body-mass-fit}
\end{figure}

We first describe  briefly the previous analysis of ${\bar B}^0\to \Lambda_c^+{\bar p}\pi^+\pi^{-}$~\cite{park},
We select $\bar{B}^0\to\Lambda_c^+\bar{p}\pi^{+}\pi^-$ events by reconstructing $\Lambda_c^+\to p K^- \pi^+$ decays,
using charged tracks reconstructed by the SVD and CDC, and hadron identification information
(such as protons, kaons and pions) provided from the CDC $dE/dx$, TOF and ACC  
(PID)~\cite{belle-pid}, and ECL and KLM information to veto electron and $\mu$ tracks.
Charge-conjugate modes are implicitly included throughout this paper unless noted otherwise.
After the event selection,
we fit the $\Delta{E}$ distribution 
for the $B$ candidate events with $5.27\,{\rm GeV}/c^2<M_{\rm bc}<5.29\,{\rm GeV}/c^2$, 
with a double Gaussian fixed to the signal MC shape
($\sigma_{\rm core}=7\,{\rm MeV}/c^2$, $\sigma_{\rm tail}=16\,{\rm MeV}/c^2$) 
plus a linear background. 
The variable $\Delta{E}=E_{B}-E_{\rm{beam}}$ is 
the difference between the reconstructed $B$ meson energy ($E_{B}$) and
the beam energy ($E_{\rm beam}$) evaluated in the center-of-mass system (CMS),
while $M_{\rm bc}=\sqrt{E_{\rm beam}^2 - P_B^2}$
is the beam-energy-constrained $B$ meson mass and
${P_B}$ is the momentum of the $B$ meson also evaluated in the CMS. 
We obtain a $B$ signal of $1400\pm49$ events for ${\bar B}^0\to \Lambda_c^+{\bar p}\pi^+\pi^{-}$.
Figure~\ref{fig:4-body-mass-fit} shows the
$\Lambda_c^+\pi^+$ and $\Lambda_c^+\pi^-$ mass distributions for the events
in the $B$ signal region $|\Delta{E}|<0.03{\,\rm GeV}$ ($\pm4\sigma$) and $5.27{\,\rm GeV}/c^2<M_{\rm bc}<5.29{\,\rm GeV}/c^2$ ($\pm4\sigma$).
We focus our discussion on the $\Sigma_c(2455)^{++}$ and $\Sigma_c(2455)^{0}$ resonances 
clearly observed in figure~\cite{park};
$(182\pm15)$ events for $\bar{B}^0\rightarrow\Sigma_{\lowercase{c}}(2455)^{++}\lowercase{\bar{p}}\pi^{-}$ 
and  $(122\pm14)$ events for $\bar{B}^0\rightarrow\Sigma_{\lowercase{c}}(2455)^{0}\lowercase{\bar{p}}\pi^{+}$,
corresponding to branching fractions of ({\nbrlcpiapp})$\times10^{-4}$ and ({\nbrlcpiazr})$\times10^{-4}$, respectively.
Hereafter, we denote $\Sigma_{\lowercase{c}}(2455)$ as $\Sigma_{\lowercase{c}}$.

\begin{figure}[!htb]
\centering
\includegraphics[width=0.4\textwidth]{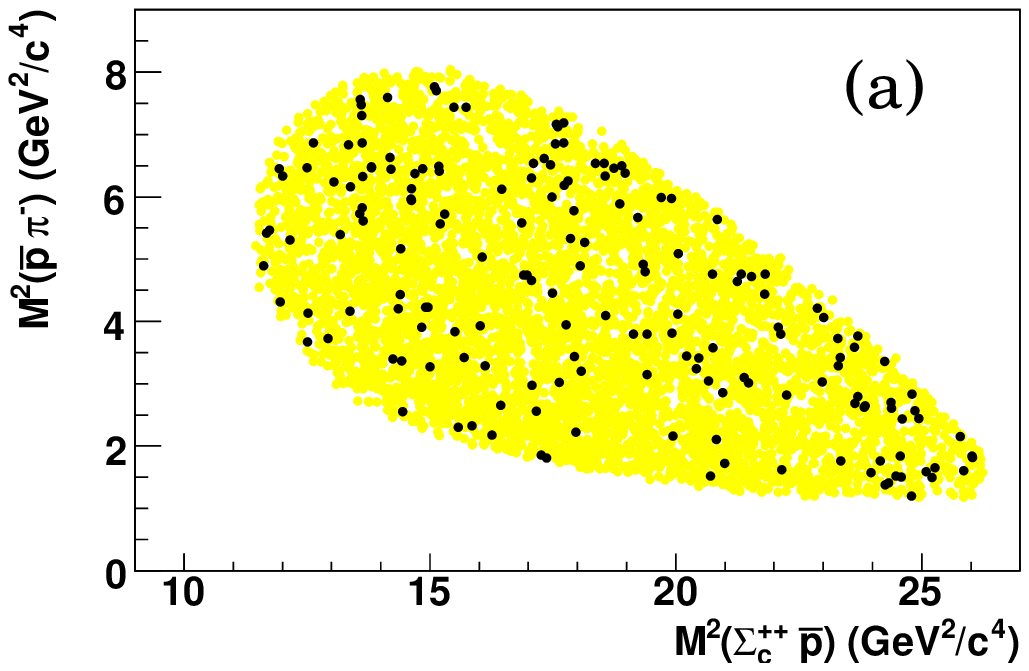}
\includegraphics[width=0.4\textwidth]{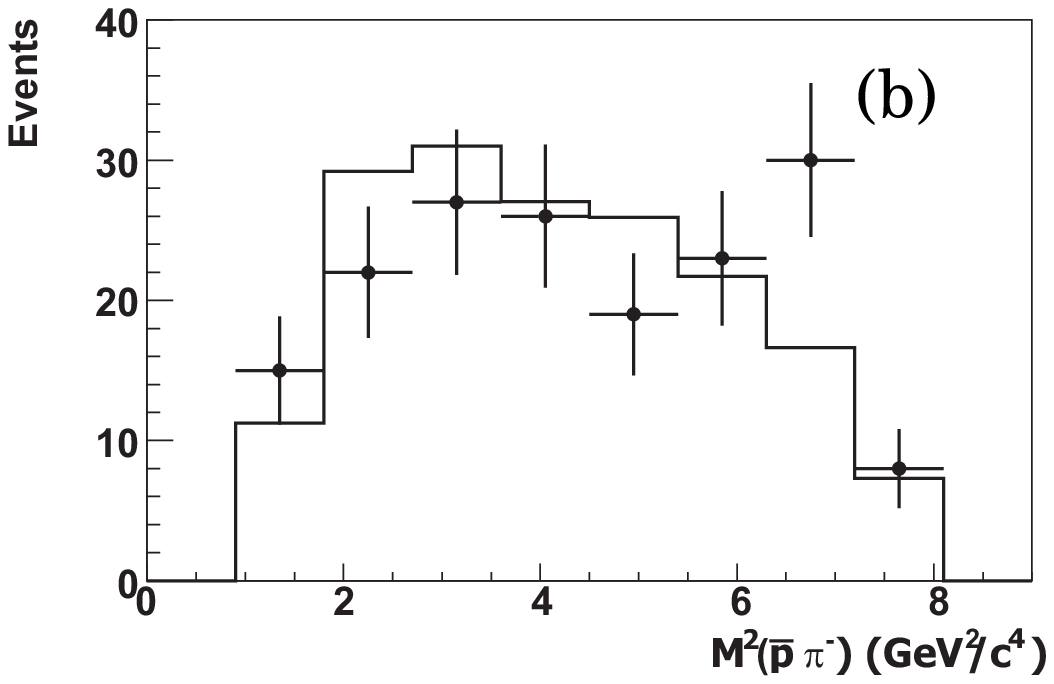}
\includegraphics[width=0.4\textwidth]{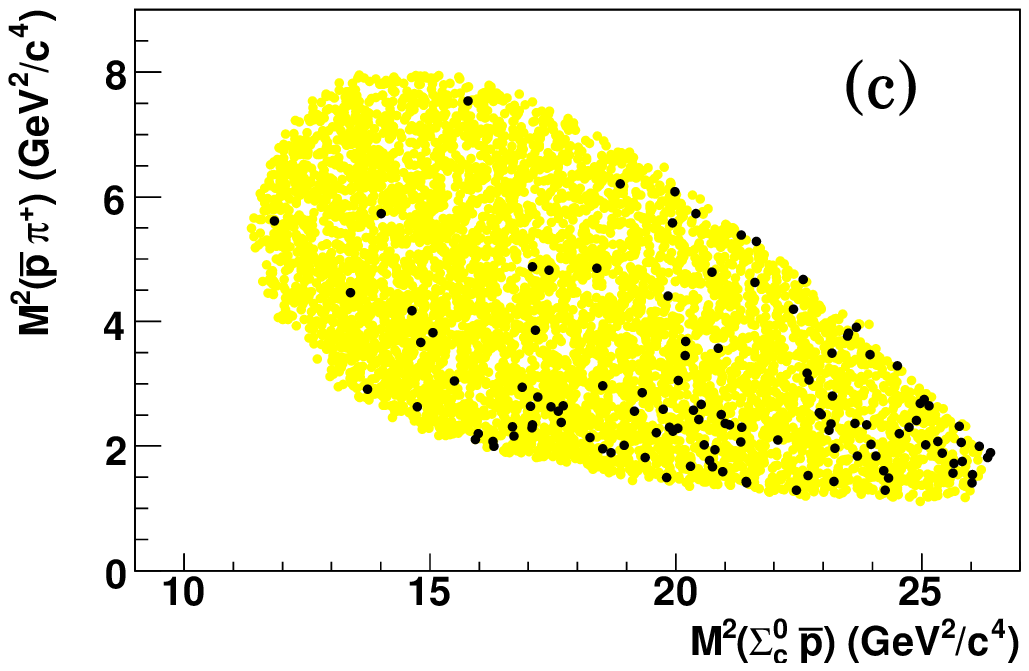}
\includegraphics[width=0.4\textwidth]{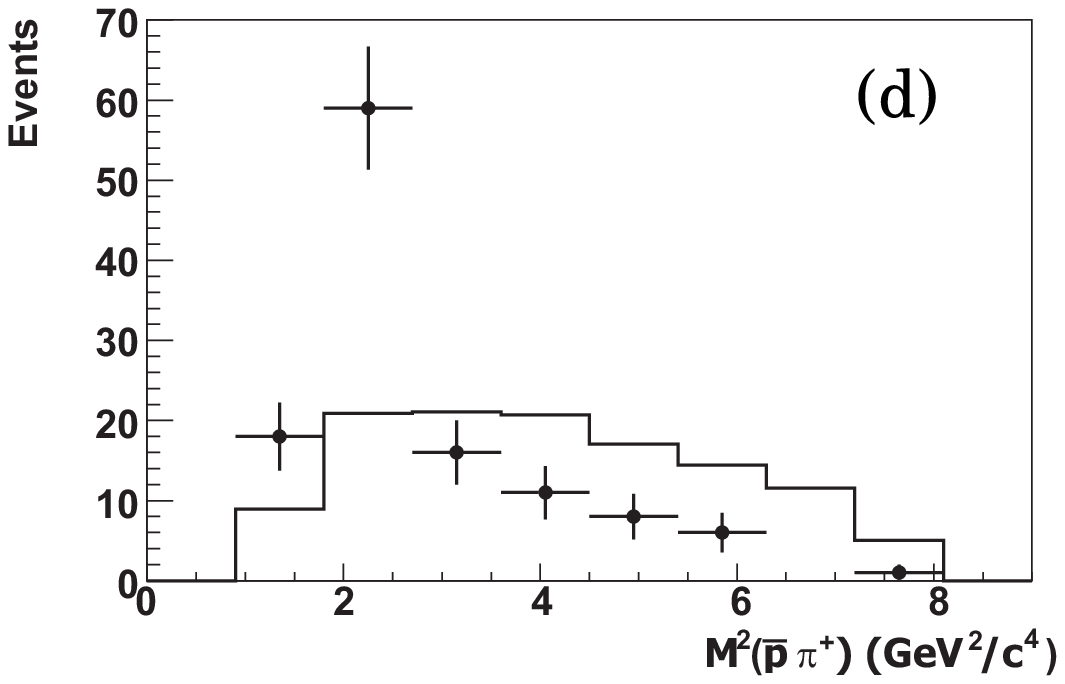}
\centering
\caption{
(a) Dalitz plot 
and (b) $M^2(\bar{p}\pi^-)$ distribution for $\bar{B}^{0}\to\Sigma_c^{++}\bar{p}\pi^{-}$.
(c) Dalitz plot 
and (d) $M^2(\bar{p}\pi^+)$ distribution for $\bar{B}^{0}\to\Sigma_c^{0}\bar{p}\pi^{+}$.
Points with error bars indicate the data, and histograms are the
decays simulated according to three-body phase space.
}
\label{fig:dalitz_sc2455}
\end{figure}

Figure~\ref{fig:dalitz_sc2455} shows (a) the Dalitz plot 
and (b) the $M^2(\bar{p}\pi^-)$ distribution for the $\bar{B}^{0}\to\Sigma_c^{++}\bar{p}\pi^{-}$ events, and (c) the Dalitz plot and 
(d) the $M^2(\bar{p}\pi^+)$ distribution for the $\bar{B}^0\to\Sigma_{c}^{0}\bar{p}\pi^+$ events.
Here we require the $\Sigma_c$ candidates satisfy the invariant mass requirement
$2.447\,{\rm GeV}/c^2<M(\Lambda_c^+\pi^{\pm})<2.461\,{\rm GeV}/c^2$ ($\pm2\sigma$).
We find that the $M^2(\bar{p}\pi^-)$ distribution for $\bar{B}^0\rightarrow\Sigma_{c}^{++}{\bar{p}}\pi^{-}$ 
is consistent with three-body phase space, while
the $M^2(\bar{p}\pi^+)$ distribution for $\bar{B}^0\to\Sigma_{c}^{0}\bar{p}\pi^+$ has a significant peak. 
In what follows, we present a detailed study of 
the $\bar{B}^0\to\Sigma_c^{0}\bar{p}\pi^{+}$ decay.

\begin{figure}[!htb]
\centering
\includegraphics[width=0.45\textwidth]{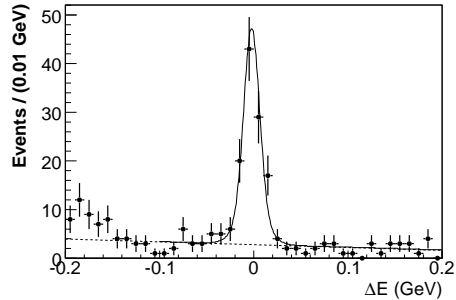}
\centering
\caption{
$\Delta{E}$ distribution for the $\bar{B}^0\to\Lambda_c^+\bar{p}\pi^+\pi^-$ events in the $M_{\rm bc}$ signal region
with $2.447\,\rm{GeV}/c^2<M(\Lambda_c^+\pi^{\pm})<2.461\,\rm{GeV}/c^2$.
The curves indicate the fit with a double Gaussian for the signal and  a linear background. 
}
\label{defit-sc2455}
\end{figure}

Figure~\ref{defit-sc2455} shows the $\Delta{E}$ distribution for the $\bar{B}^0\to\Sigma_c^{0}\bar{p}\pi^{+}$
events, which are selected from the $\bar{B}^0\to\Lambda_{c}^{+}\bar{p}\pi^+\pi^-$ sample
with the additional requirement that the $\Lambda_c^+\pi^-$ mass be consistent with the $\Sigma_c^0$.
The curves show fits to the data with a double Gaussian function with shape parameters
fixed to the values from signal MC and a linear background. 
We obtain a $B$ signal yield of $(102\pm11)$ events 
and a background of $(17\pm3)$ events.
The signal reduction of 16\% is consistent with the MC estimation of the effect
due to the ${\Sigma_c^0}$ mass requirement.
We estimate a non-$\Sigma_c^0$ background of $(8\pm4)$ events 
from a fit to the $\Delta{E}$ distribution in the $\Sigma_c$ mass sideband
$2.435{\,\rm GeV}/c^2<M(\Lambda_c^+\pi^{-})<2.442{\,\rm GeV}/c^2$ and 
$2.466{\,\rm GeV}/c^2<M(\Lambda_c^+\pi^{-})<2.473{\,\rm GeV}/c^2$. 
This can be compared with $(2.5\pm0.5)$ events estimated from MC simulation of
$\bar{B}^0 \rightarrow \Lambda_{c}^{+} \bar{p} \pi^{+} \pi^{-}$ decay
with four-body phase space normalized to the total of 1400 events~\cite{park}.
Here the error is due to the statistics of the simulation.
We do not take into account the non-$\Sigma_c^0$ background in the analysis that follows.

\begin{figure}[!htb]
\centering
\includegraphics[width=0.32\textwidth]{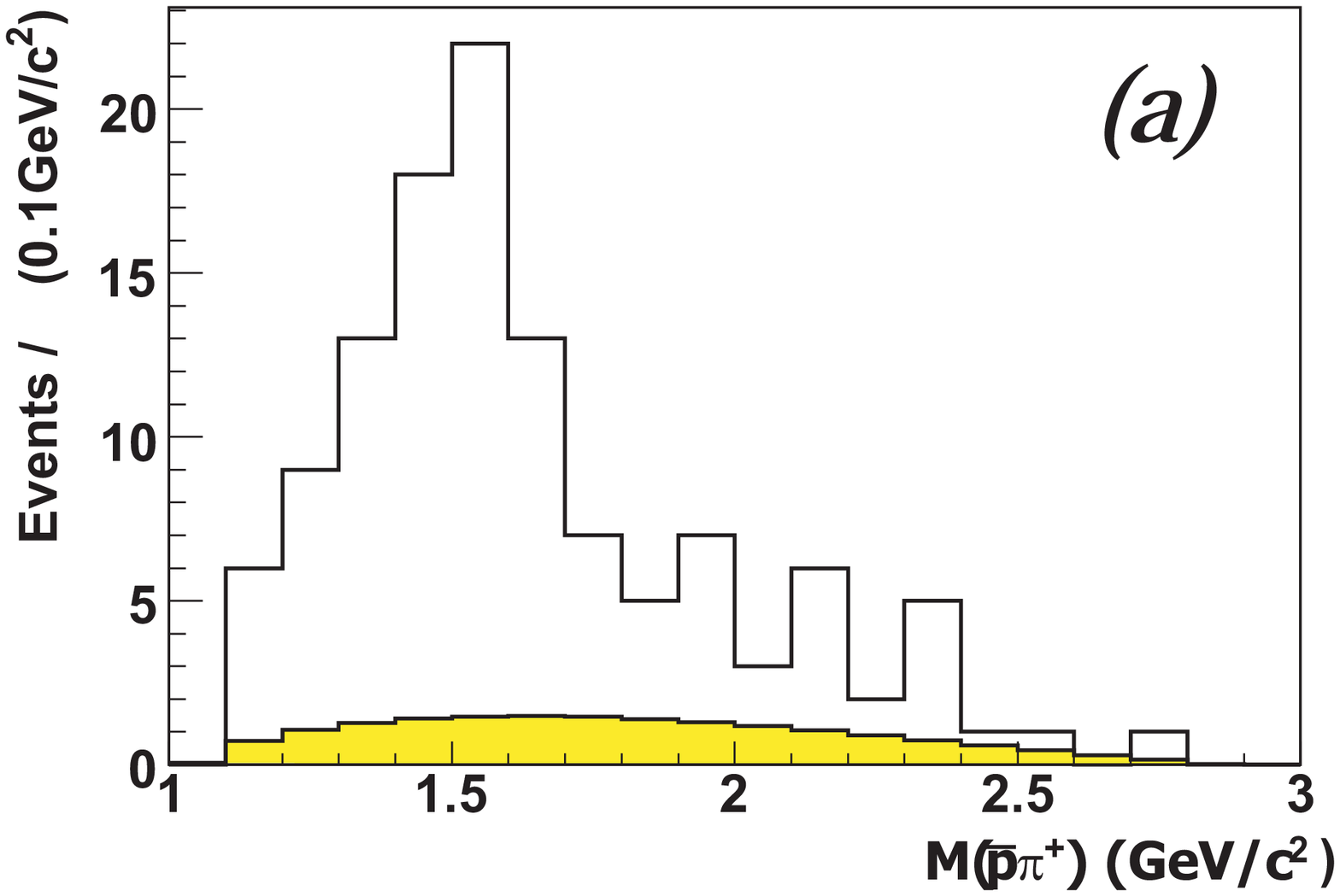}
\includegraphics[width=0.32\textwidth]{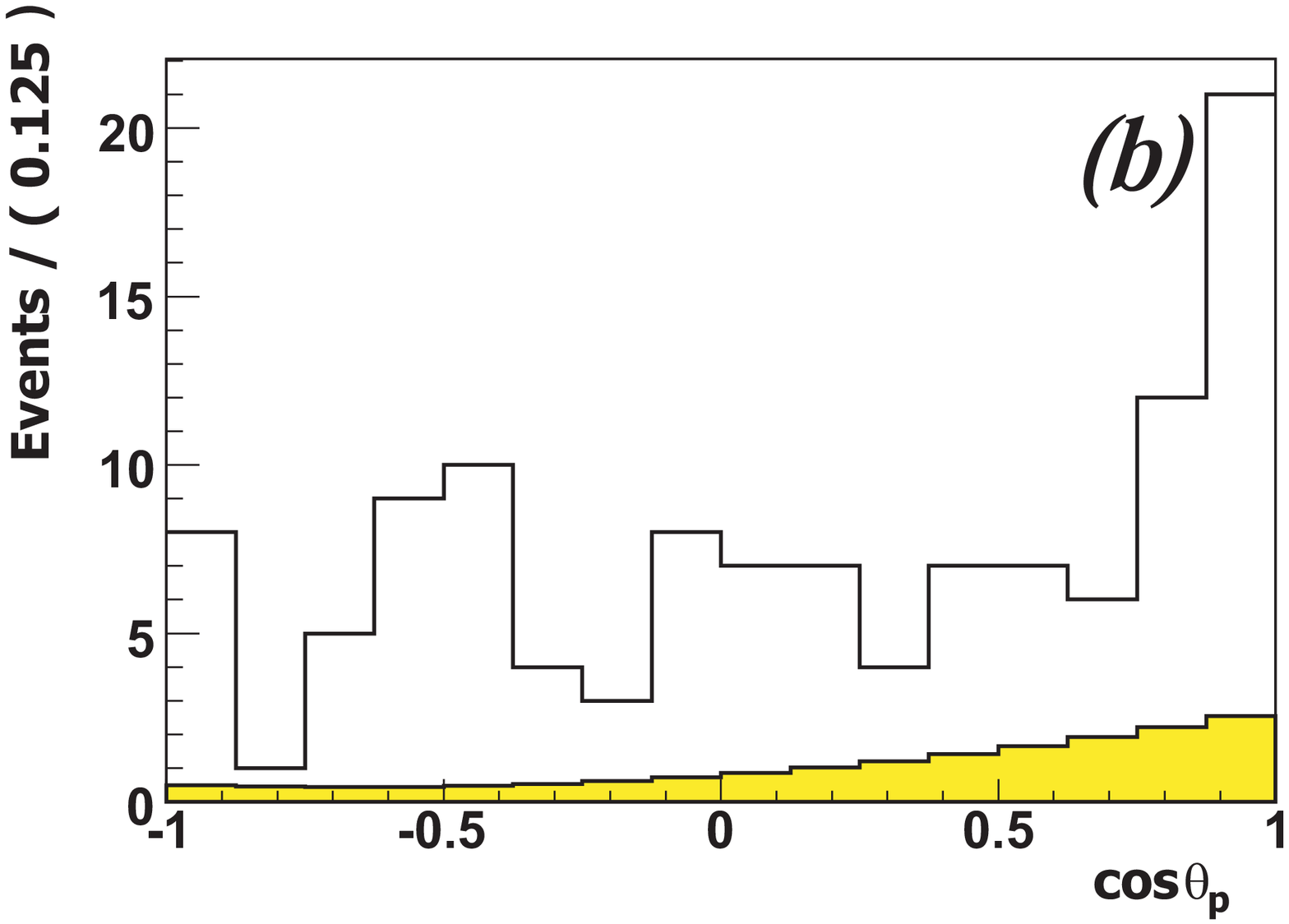}
\includegraphics[width=0.32\textwidth]{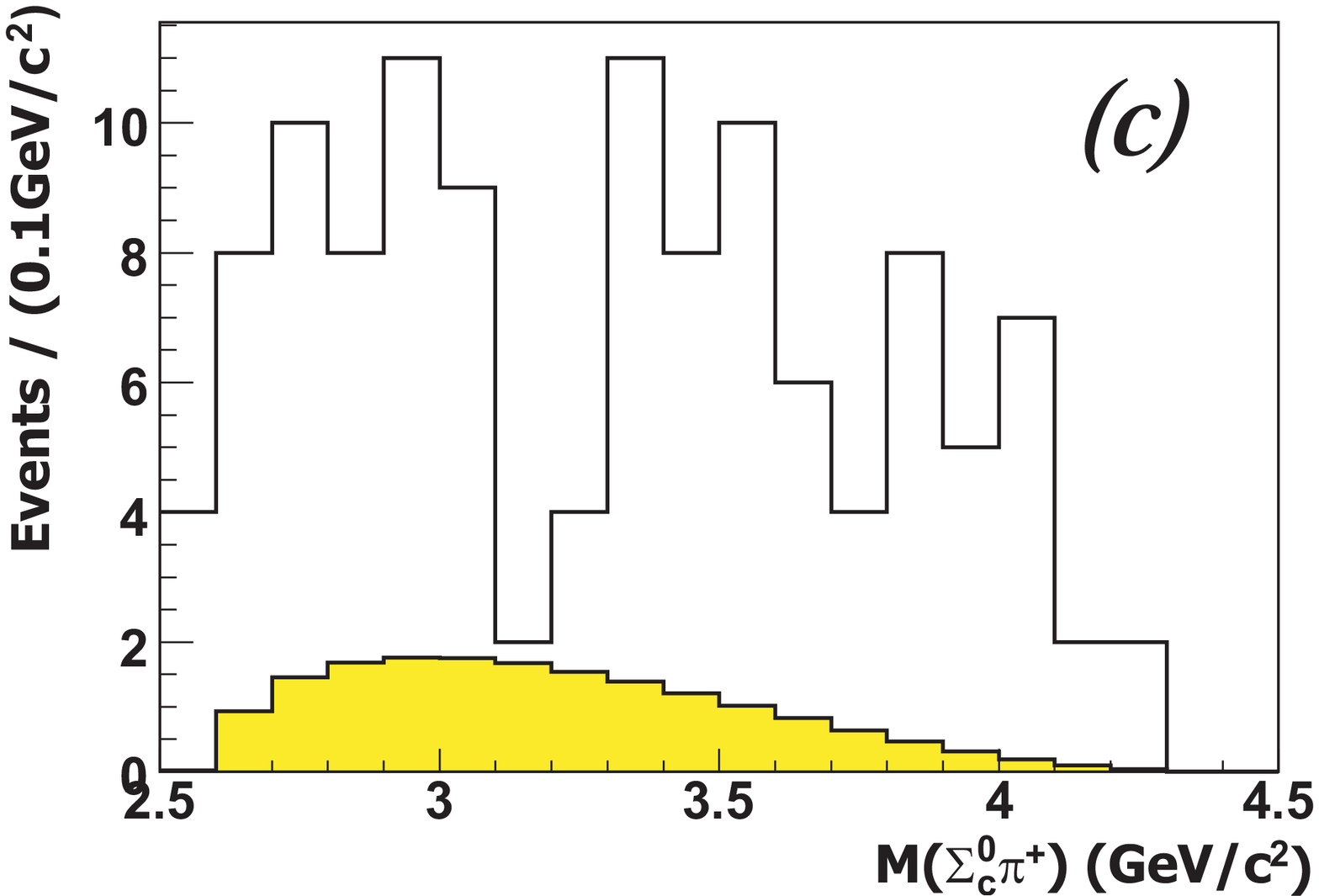}
\centering
\caption{
Data distributions for
(a) $M(\bar{p}\pi^+)$, (b) $\cos\theta_p$ and (c) $M(\Sigma_c^0\pi^+)$. 
The shaded histograms indicate the normalized background.
}
\label{data-3hist}
\end{figure}

Figure~\ref{data-3hist} shows (a) the $\bar{p}\pi^+$ mass, (b) $\cos\theta_{p}$ and 
(c) $\Sigma_c^0\pi^+$ mass distributions for the selected {\thbody} events. 
Here, $\cos\theta_{p}$ is the cosine of the angle between the $\bar{p}$ momentum 
and the direction opposite to the $B$ momentum in the $\bar{p}\pi^+$ rest frame.
The shaded histograms indicate the distributions for the background discussed above.
The background shapes are obtained by fits to the data in the sideband region  
$|\Delta{E}|<0.1\,{\rm GeV}$ and $5.26\,{\rm GeV}/c^2<M_{\rm bc}<5.29\,{\rm GeV}/c^2$
outside the $B$ signal region, and the yield is fixed to 17 events.
Here, the $M(\bar{p}\pi^+)$ distribution is parameterized by the function
$P_{\rm bkg}(M)=c\cdot\sqrt{t_{\rm min}\cdot t_{\rm max}}\cdot (1+c_1\cdot t_{\rm min})(1+c_2\cdot t_{\rm max})$
with $t_{\rm min}=(M^2-M^2_{\rm min})$ and $t_{\rm max}=(M^2_{\rm max}-M^2)$.
$M_{\rm min}$ and $M_{\rm max}$ are the minimum and maximum masses.
The variable $c$ is a normalization constant, 
and $c_1$ and $c_2$ are shape parameters.
The $\cos\theta_p$ distribution is modeled by a second-order Chebyshev polynomial.
 
We find a significant structure in the $\bar{p}\pi^+$ mass distribution,
and a forward peak in the $\cos\theta_{p}$ distribution, and a low mass $\Sigma_c^0\pi^+$ enhancement, denoted as {\lcx}. 
The $\bar{p}\pi^+$ mass structure has a mass near $1.5\,{\rm GeV}/c^2$ and a width of about $0.3\,{\rm GeV}$.
We denote it as $\bar{N}^0$, and investigate its characteristics in detail.
In order to describe the $\bar{p}\pi^+$ mass structure, 
which is not explained by a simple phase space non-resonant {\MCB} decay,
we consider an intermediate two-body decay {\MCA} with a 
resonant state ${\bar N}^0\to{\bar p}\pi^+$.
However, we still cannot reproduce 
the forward $\cos\theta_{p}$ peak and the $\Sigma_c^0\pi^+$ low mass structure
with these two modes only.
Therefore, we introduce one additional mode {\MCC} to account for the observed features.
As the low mass $\Sigma_c^0\pi^+$ structure is close to threshold, it produces
a forward peak in the $\cos\theta_{p}$ distribution.
In the low $\Sigma_c^0\pi^+$ mass region, 
we search for known $\Sigma_c^0\pi^+$ resonant states~\cite{pdg2006} in finer mass bins,
but find no signals. So far, there is no good candidate to interpret this broad structure as a resonance.
Therefore we assume that there is a threshold mass enhancement with
a mass of $2800\,{\rm MeV/}c^2$ and a width of $350\,{\rm MeV}$ obtained from a fit to the $\Sigma_c^{0}\pi^+$
mass distribution using a relativistic Breit-Wigner (S-wave) function.

\begin{figure}[!htb]
\centering
\includegraphics[width=0.32\textwidth]{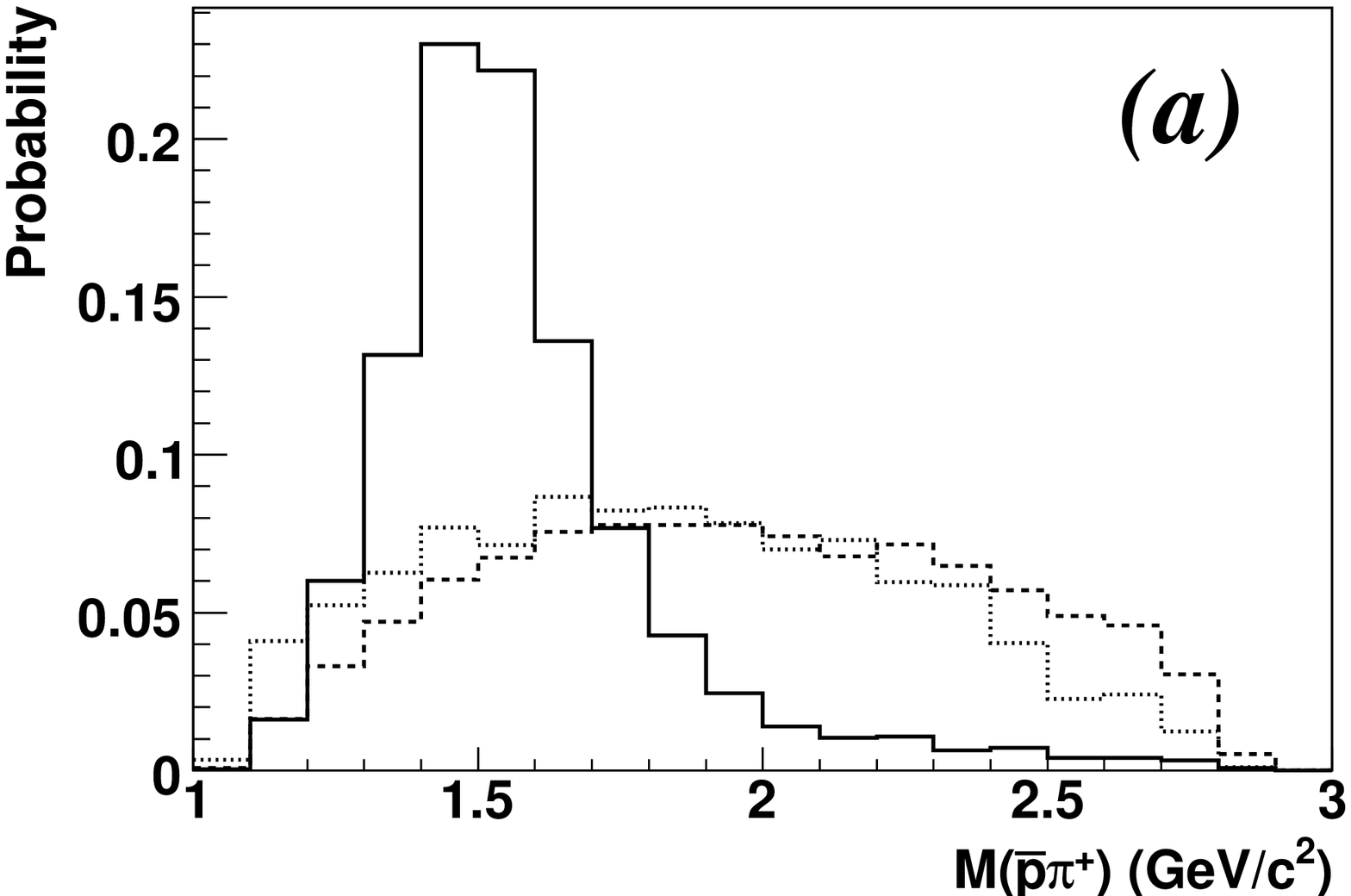}
\includegraphics[width=0.32\textwidth]{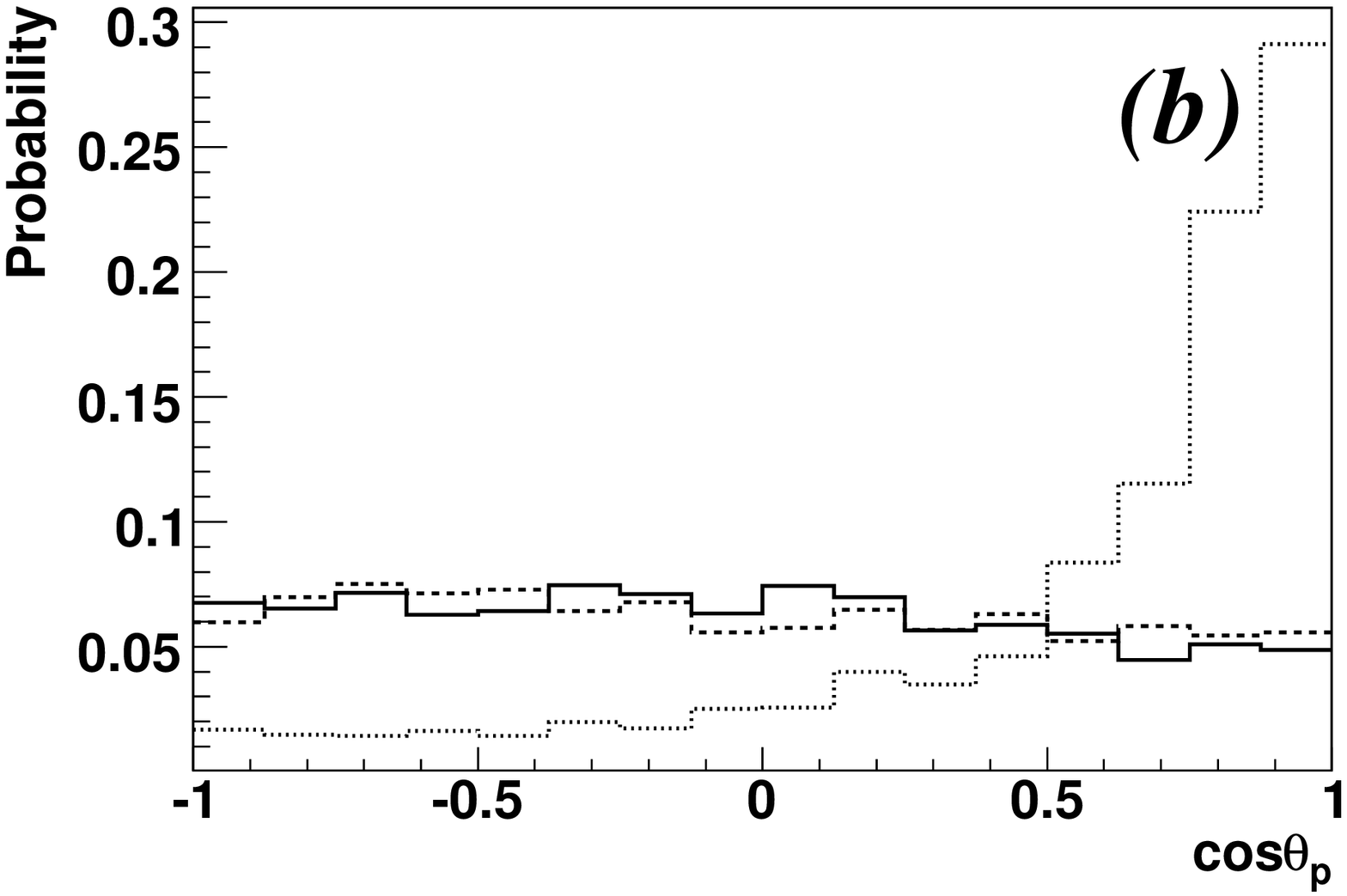}
\includegraphics[width=0.32\textwidth]{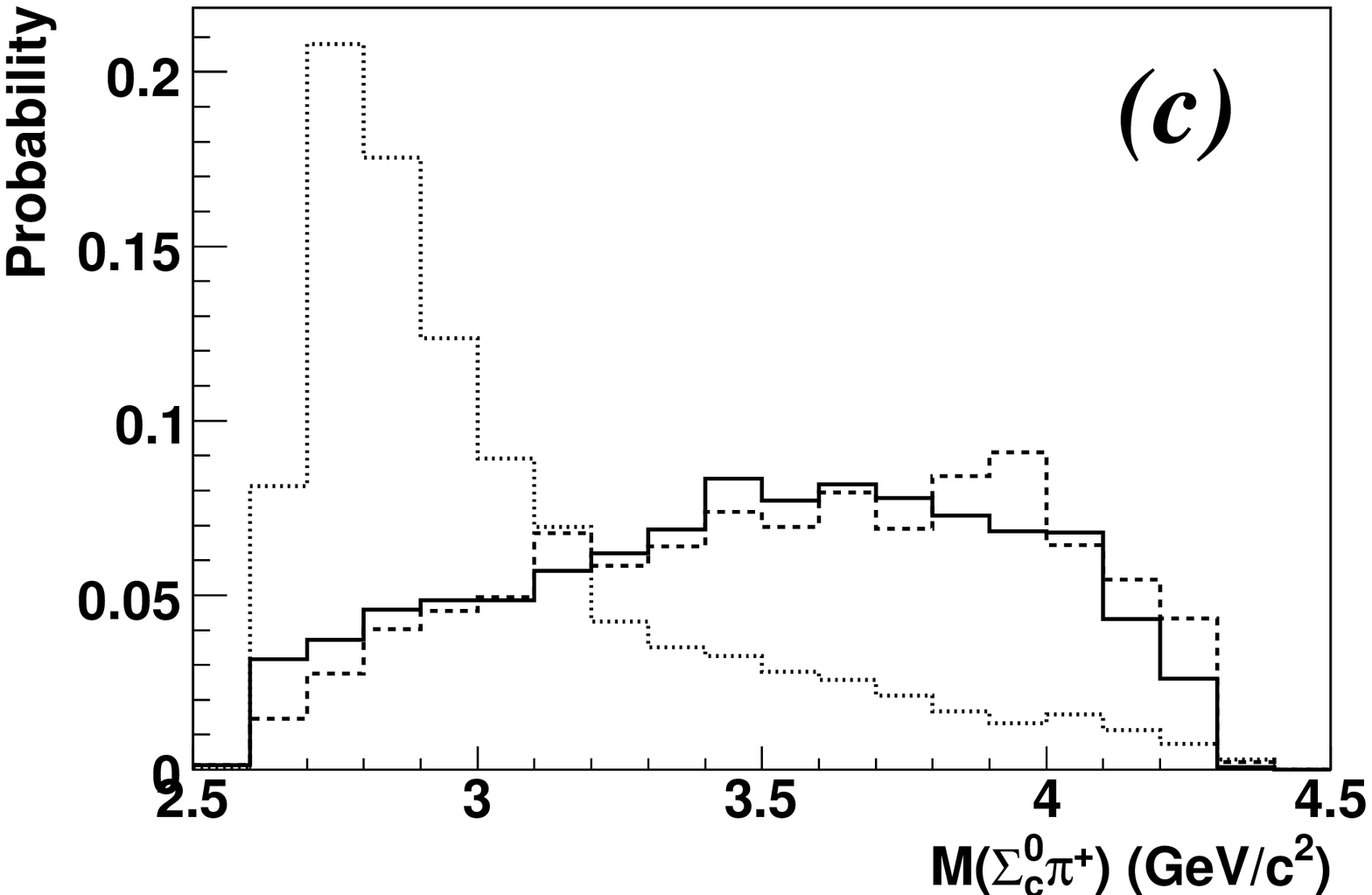}
\centering
\caption{
Binned probability distributions of (a) $M(\bar{p}\pi^+)$, (b) $\cos\theta_p$ and (c) $M(\Sigma_c^0\pi^+)$,
where we compare MC simulated distributions
for {\MCA} (solid lines), {\MCB} (dashed lines), and {\MCC} (dotted lines).
We make a simultaneous fit to 
the distributions in (a) and (b).
 }
\label{mc-3hist}
\end{figure}

Figure~\ref{mc-3hist} compares binned Probability Density Functions (PDF) of
the MC simulated events for the three assumed decay modes.
The histograms show the PDFs for
(a) the $M(\bar{p}\pi^+)$, (b) $\cos\theta_p$ and (c) the $M(\Sigma_c^0\pi^+)$ distributions.
The solid histograms show the distributions of the mode {\MCA},
assuming a P-wave relativistic Breit-Wigner amplitude with a mass of $1530\,{\rm MeV/}c^2$ 
and a width of $340\,{\rm MeV}$. 

To determine the $N^0$ mass, width and the yields of the three modes,
we perform a maximum likelihood fit to
the observed $M({\bar p}\pi^+)$ and $\cos\theta_p$ distributions shown in Fig.~\ref{data-3hist}.
These two distributions are sufficient to fully describe the three-body decay {\MCB}.
To model the observed distribution, we construct a function $F(M(\bar{p}\pi^+),\cos\theta_p)$
from the sum of PDFs of the three decay modes and the background.
\begin{equation*}
\begin{split}
F=\nu_1 P_{\bar{B}^0\to\Sigma_{c}^{0}\bar{N}^0}
+\nu_2 P_{\bar{B}^0\to(\Sigma_{c}^{0}\pi^+)_{\rm X}\,\bar{p}} \\
+\nu_3 P_{\bar{B}^0\to\Sigma_{c}^{0}\bar{p}\pi^+}
+\nu_4 P_{\rm bkg},
\end{split} 
\end{equation*}
where $P_i$ denotes a product of the normalized PDFs,
$Q_i(M(\bar{p}\pi^+))$ (20 bins) and $R_i(\cos\theta_p)$ (16 bins),
and $\nu_i$ stands for the yield of the $i$-th mode.
We plot $Q(M(\bar{p}\pi^+))$ and $R(\cos\theta_p)$ distributions from the detector MC simulation 
for {\MCC} and {\MCB} modes, as shown in Figs~\ref{mc-3hist}(a) and (b), respectively.
For the {\MCA} mode, we use the MC simulated $R(\cos\theta_p)$ distribution and 
a Breit-Wigner for $Q(M({\bar p}\pi^+))$ with the ${\bar N}^0$ mass and width ($m_R$, $\Gamma$) as free parameters.
A small systematic error due to the use of the $Q(M({\bar p}\pi^+))$ distribution without the MC detector simulation 
is discussed later.
We use a P-wave (S-wave) relativistic Breit-Wigner shape.
\begin{equation*}
BW_P(m^2)=\frac{ p^2 } 
{ (m^2-m_R^2)^2 + m_R^2{\Gamma^2(m)}} 
{\left[ \frac{B(p)}{B(p_0)}\right]}^2,
\end{equation*}
\begin{equation*}
BW_S(m^2)=\frac{m_R\Gamma(m)}{[(m^2-{m_R}^2)^2+{m_R}^2\Gamma^2(m)]}, 
\end{equation*}
\begin{equation*}
\Gamma(m)=\left( \frac{p}{p_0}\right)^{2L+1}
\left( \frac{m_R}{m}\right) \Gamma_0
{\left[ \frac{B(p)}{B(p_0)} \right]^2}.
\end{equation*}
Here $m$ is the mass of the $\bar{p}\pi^+$ system, and $m_R$ is the nominal $\bar{N}^0$ mass, and $\Gamma(m)$ is the width.
The variable $p$ is the momentum of a daughter particle in the $\bar{N}^0$ rest frame, and $p_0$ is that for the nominal $\bar{N}^0$ mass.
$B(p)=1/{\sqrt{1+(Rp)^2}}$ is the Blatt-Weisskopf form factor~\cite{blatt}.
The value $R$, called the centrifugal barrier penetration factor, is set to $3\,{\rm (GeV}/c)^{-1}$ for a P-wave, and
is zero for an S-wave. $L$ indicates the orbital angular momentum.
For the S-wave Breit-Wigner amplitude~\cite{pilkuhn} we use $\Gamma(m)$ with $m=m_R$ and $L=0$
to parameterize the smooth shape near the mass threshold.

We define an extended unbinned likelihood with coarse bins, 
and carry out a maximum likelihood fit.
\begin{equation*}
L=\frac{e^{-(\nu_1+\nu_2+\nu_3+\nu_4)}}{N!}\prod F(m_R,\Gamma,\nu_1,\nu_2,\nu_3,\nu_4)
\end{equation*}
 We fit the ${\bar N}^0$ mass and width $(m_R,\Gamma)$ 
and the yields $\nu_1, \nu_2$ and $\nu_3$ as free parameters, while  the background yield 
$\nu_4$ is fixed to 17 events.
Table~\ref{fit-detail2} summarizes the fit results with various model assumptions. 
We calculate the statistical significance from the quality -2${\rm ln}({L}_{0}/{L}_{\rm max})$, where
$L_{\max}$ is the maximum likelihood returned from the fit, and $L_{0}$
is the likelihood with the signal yield fixed to zero, 
and taking into account the reduction of the degrees of freedom.
We obtain a significance of {\sgnfn}$\,\sigma$ for the {\MCA} contribution. 
The signal in the mode {\MCC} has a statistical significance of {\sgnfl}$\,\sigma$,
while that for {\MCB} is not significant (0.8\,$\sigma$). 
Here, we calculate the goodness-of-fit from the likelihood ratio $\lambda$~\cite{pdg2006},
\begin{equation*}
\begin{split}
\chi^2 \approx -2{\rm ln} \lambda 
=2\sum_{j=1}^{36}[{\cal F}_j(m_R,\Gamma,\nu_1,\nu_2,\nu_3,\nu_4)-F_j  \\
+F_j{\rm ln}(\frac{F_j}{{\cal F}_j(m_R,\Gamma,\nu_1,\nu_2,\nu_3,\nu_4)})],
\end{split}
\end{equation*}
where $F_j$ and ${\cal F}_j$ are the observed and the fitted yields, respectively, in the $j$-th bin:
$j=1, 20$ for 20 bins in $Q(M(\bar{p}\pi^+))$ and $j=21, 36$ for 16 bins in $R(cos\theta_p)$.

The small contribution ($\nu_3=-11\pm10$) can be understood from Fig.~\ref{mc-3hist}.
The mode {\MCB} has a broad $\bar{p}\pi^+$ mass distribution similar to {\MCC},
while it does not reproduce the forward $\cos\theta_{\bar p}$ peak.
On the other hand, the {\MCA} mode can reproduce the ${\bar p}\pi^+$ mass bump structure and
the uniform  $\cos\theta_{\bar p}$ distribution. 
Hence, we fix $\nu_3=0$ in the subsequent fit and
the uncertainty of this contribution is taken into account in the systematic error. 

\begin{table}[!htb]
\caption{ 
Summary of the simultaneous fits to the 
$M(\bar{p}\pi^+)$ and $\cos\theta_{\bar p}$ distributions
with the three decay modes {\MCA}, {\MCB} and {\MCC}.
(a) - (d) represent fits with various assumed contributions.
Here, we show the fit results with the P-wave assumption, as we find
no significant difference from the S-wave assumption.
}
\begin{center}
\begin{tabular}{c|cccccccc} \hline
Decay mode       & (a)   & (b)   & (c)   & (d)  & Signif. \\ \hline
{\MCA}           & free  & free  & free  & 0    & 7.0 \\ \hline
{\MCC}           & free  & free  & 0     & free & 4.6 \\ \hline 
{\MCB}           & free  & 0     & free  & free & 0.8 \\ \hline \hline
$\chi^2$/ndf & 31.7/31  & ~32.4/32  & ~52.8/32  & ~88.4/34 &  & \\ \hline 
\end{tabular}
\label{fit-detail2}
\end{center}
\end{table}

\begin{figure}[!htb]
\centering
\includegraphics[width=0.62\textwidth]{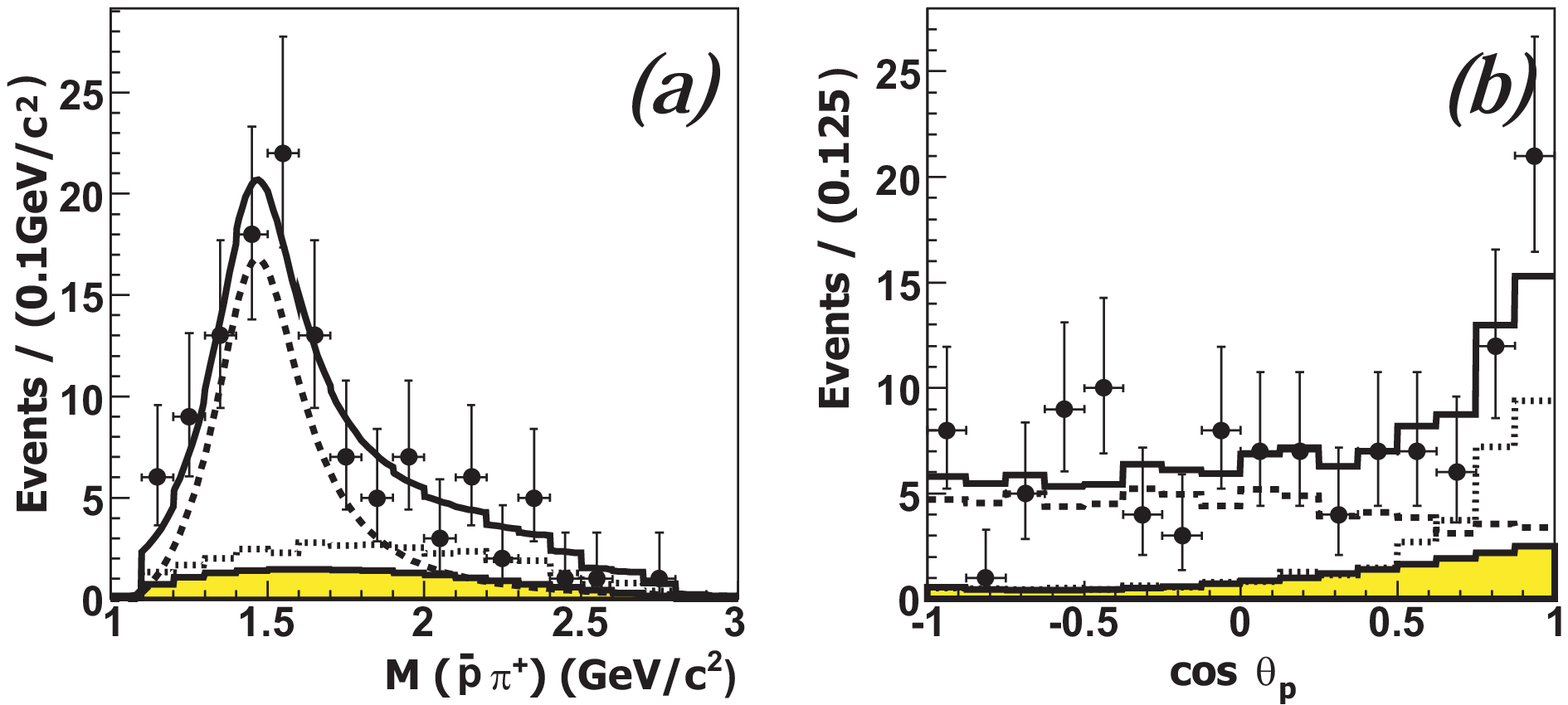}
\includegraphics[width=0.31\textwidth]{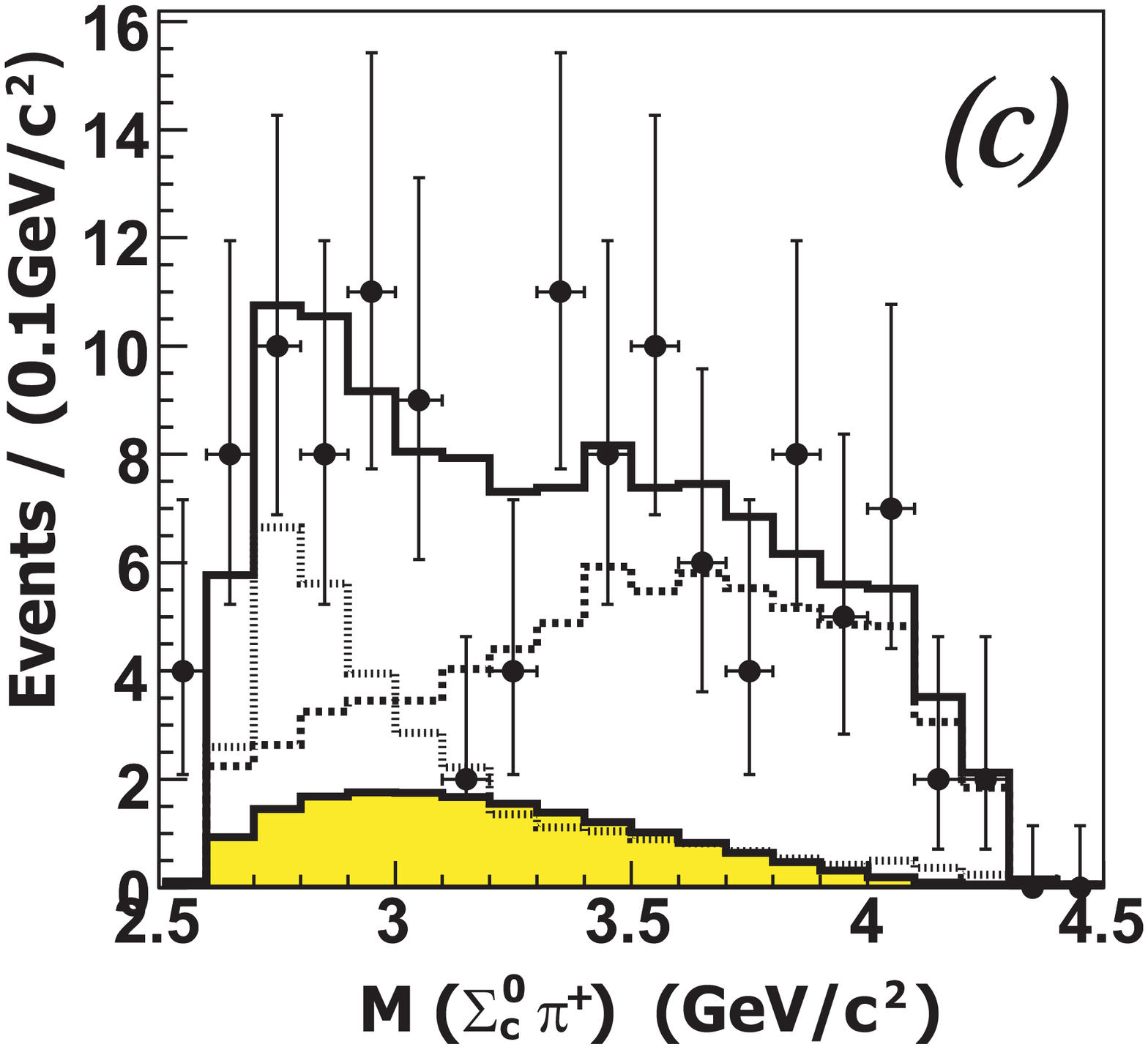}
\centering
\caption{
Simultaneous fit to (a) $M({\bar p}\pi^+)$ and (b) $\cos\theta_p$ distributions with a P-wave Breit-Wigner.
The points with error bars are the data, and
the curves are the contributions from {\MCA} (dashed), {\MCC} (dotted), the background (shaded)
and their sum (solid).
(c) $M(\Sigma_c^0\pi^+)$ distribution, where the curves represent
their contributions obtained by the fit to (a) and (b).
}
\label{sim-fit}
\end{figure}

Figure~\ref{sim-fit} shows the results of a fit to (a) the $M({\bar p}\pi^+)$ 
and (b) $\cos\theta_{\bar p}$ distributions
under the P-wave assumption.
The data are the points with error bars.
The curves are the contributions from {\MCA} (dashed), {\MCC} (dotted), the background (shaded)
and their sum (solid).
We obtain yields of ({\yieldp}) and ({\yieldl}) for the modes {\MCA} and {\MCC}, respectively. 
Figure~\ref{sim-fit}(c) shows that 
the $M(\Sigma_c^0\pi^+)$ distribution is consistently represented by the fitted parameters
even though the $M(\Sigma_c^0\pi^+)$ distribution is not included in the fit.

\begin{table}[!htb]
\caption{
The fitted ${\bar N}^0$ mass and width with relativistic S-wave and P-wave Breit-Wigners.
The first errors are statistical and the second are systematic including
the uncertainties in the yields of {\MCB} and the background, and the assumption of
a low mass {\lcx} structure.
}
\begin{center}
\begin{tabular}{c|cccc} \hline
 Item  & Yield & Mass & $\Gamma$ & $\chi^2$/ndf \\ \hline 
       & Events & $\rm{MeV}/c^2$ & $\rm{MeV}$ &  \\ \hline 
S-wave  & \yieldss & ~\masss & ~\widths & ~32.9/32 \\ \hline
P-wave  & \yieldsp & ~\massp & ~\widthp & ~32.4/32 \\ \hline
$\Sigma_c^0\pi^+$ sys.   &  $\pm5$   & $\pm8$ & $\pm50$ &  \\ \hline
\end{tabular}
\label{bw-fit-systematics}
\end{center}
\end{table}

Table~\ref{bw-fit-systematics} compares the fit results for the ${\bar N}^0$ yield, 
mass and width with P-wave and S-wave assumptions.
The fitted yields are found to be comparable with each other, while the mass and width show a systematic difference. 
We estimate systematic errors by varying the fitted yields by $\pm\sigma$ for
the background ($\pm3$) and {\MCB} ($\pm11$), and by taking into account the
uncertainty in modeling the low mass {\lcx} structure as discussed in the following.  

The simulated $R(\cos\theta_p)$ distribution for {\MCA} is almost flat, 
as the generated $\cos\theta_p$ distribution is uniform for $P$- and $S$-waves.
However, the $R(cos\theta_p)$ distribution is slightly affected 
by the assumed BW parameters due to efficiency changes in $\cos\theta_p$. 
We study the systematics of the fitted ${\bar N}^0$ mass and width
due to the assumption on $R(\cos\theta_p)$,
by changing the mass and the width in EvtGen 
in ranges between $1400\,{\rm MeV}/c^2$ and $1620\,{\rm MeV}/c^2$, 
and between $200\,{\rm MeV}$ and $450\,{\rm MeV}$, respectively.
We find variations of $\pm1\,{\rm MeV}/c^2$ in the fitted mass 
and $\pm5\,{\rm MeV}/c^2$ in the width.
We also study the systematic errors due to the parameterization
of the $(\Sigma_c^0\pi^+)$ low mass structure.
Instead of assuming a model with a single Breit-Wigner {\lcx},
we consider a combination of known states $\Lambda_c^*\to\Sigma_c(2455)^0\pi^+$;
$\Lambda_c^+(2625)$ ($\Gamma_{\rm total}<1.9\,{\rm MeV}$)~\cite{pdg2006}, 
$\Lambda_c^+(2765)\to\Lambda_c^+\pi^+\pi^-$ ($\Gamma\sim50\,{\rm MeV}/c^2$),
$\Lambda_c^+(2880)$ ($\Gamma=5.8\pm1.3\,{\rm MeV}$), 
and $\Lambda_c^+(2940)$ ($\Gamma=13^{+28}_{-9}\,{\rm MeV}$).
Here we use the partial widths for $\Sigma_c^0\pi^+$ decay of the last three states
given by Ref.~\cite{mizuk}.
We make a fit to the $\bar{N}^0$ mass, width and the yield of {\MCA} with the individual 
{\lcknown} yields floated and with the background fixed as mentioned previously.
We obtain $N({\bar p}\pi^+)$ mass and width values in good agreement 
with those obtained by the fit with the {\MCC} model. 

The branching fraction product 
${\cal B}$(\MCA)$\times{\cal B}(\bar{N}^0\to\bar{p}\pi^+)$
is calculated as $N_s/(N_{B\bar{B}} \cdot \epsilon \cdot CF \cdot {\cal B}_{\Lambda_{c}^+ \to p K^- \pi^+} )$
assuming $N_{B^{+}B^{-}}=N_{B^0\bar{B}^0}$. For $N_s$ we use the P-wave yield in Table~\ref{bw-fit-systematics} 
as it gives a better confidence level than an S-wave fit.
We use $N_{B\bar{B}} = (387.7 \pm 4.8)\times 10^6$ for the integrated luminosity of 357 fb$^{-1}$, and
the signal efficiency $\epsilon = (5.18\pm0.13$)\% from the MC simulation of {\MCA}.
We apply a correction factor $CF=(86.7\pm7.9)$\%, which takes into account the systematic difference
in particle identification (PID) between data and MC simulation. 
Correction factors for proton, kaon and pion tracks
are determined from a comparison of data and MC simulation for large
samples of $D^{*+}\to D^0(K\pi)\pi^+$ and $\Lambda\to p\pi^-$ decays. The overall PID correction factor
is then calculated as a linear sum over the six tracks for the selected $B$ signal events.
We assign an error of 7.2\% 
due to track reconstruction efficiency for the six charged tracks in the final state.
The systematic error on the branching fraction arising from a quadratic sum of the uncertainties on $N_{B\bar{B}}$,
the signal efficiency $\epsilon$, and particle identification $CF$ and track reconstruction,
is found to be 12\%. 
Including the systematic error in the {\MCA} yield,
we arrive at the total systematic uncertainty in the branching fraction of 17.6\,\%.
Thus, we obtain the branching fraction product of {\brproduct}, and
a significance of {\ssgnfn} standard deviations including systematics.
The last error is due to an uncertainty in ${\cal B}(\Lambda_{c}^+ \to p K^- \pi^+)=(5.0\pm1.3)\%$~\cite{pdg2006}.

Next, we investigate goodness-of-fits with the masses and widths fixed 
to representative values for $N({\bar p}\pi^+)$ states~\cite{pdg2006}, and
by floating the yields for {\MCA} and {\MCC}.
The fit results are summarized in Table~\ref{fit-n-states}. 
Here $L_{2I,2S}$ stands for a resonance of isospin $I$ and spin $S$ with an orbital angular momentum of 
$L$=S, P and D for $L$=0, 1 and 2, respectively.
We exclude $\Delta$ states such as $\Delta(1600)P_{33}$ and $\Delta(1620)S_{31}$, 
as we have no significant structure in the $\bar{p}\pi^-$ mass distribution in $\bar{B}^0\to\Sigma_c^{++}\bar{p}\pi^-$ decay.
The fits favor $N(1440)P_{11}$ and $N(1535)S_{11}$,
while they disfavor $N(1520)D_{13}$ and $N(1650)S_{11}$.
In the decay $\bar{B}^0\to\Sigma^0_c\bar{N}^0$ (assuming $S(\Sigma^0_c)=\frac{1}{2}$),
one expects a uniform $\cos\theta_p$ distribution for the $N(1440)P_{11}$ state, $N(1535)S_{11}$ state and 
$N(1650)S_{11}$ state, and a $(1+3\cos^2\theta)$ distribution for the $N(1520)D_{13}$ state. 
As shown in Fig.~\ref{sim-fit}(b), the distribution has a peak only in the forward direction, which
is well reproduced by the mode {\MCC}. The remaining uniform distribution is due to {\MCA}. 
Thus, the observed $\cos\theta_{\bar {p}}$ distribution  
is consistent with both, $\bar{N}(1440)P_{11}$ and $\bar{N}(1535)S_{11}$ states,
with a preference for the former due to the width of the state.

\begin{table}[!htb]
\caption{
Results of the fits using the parameters of known $N({\bar p}\pi^+)$ resonances~\cite{pdg2006}.
}
\begin{center}
\begin{tabular}{cccccccc} \hline
States &  & Mass & $\Gamma$ &  $\Sigma_c^0\bar{N}^0$ & $(\Sigma_c^0\pi^+)_X\bar{p}$  & ~$\chi^2$/ndf \\ 
    &  & ~~${\rm MeV}/c^2$ & $~{\rm MeV}$ &  ~~Events & Events  &  \\ \hline
N(1440) & $P_{11}$ & 1440 & 300 & $65\pm10$ & $39\pm9$  & 37.6/34 \\ \hline
N(1520) & $D_{13}$ & 1520 & 115 & $46\pm9$  & $53\pm10$ & 53.5/34 \\ \hline
N(1535) & $S_{11}$ & 1535 & 150 & $58\pm10$ & $43\pm10$ & 40.1/34 \\ \hline
N(1650) & $S_{11}$ & 1655 & 165 & $44\pm10$ & $55\pm11$ & 74.2/34 \\ \hline
\end{tabular}
\label{fit-n-states}
\end{center}
\end{table}

\begin{figure}[!htb]
\centering
\includegraphics[width=0.9\textwidth]{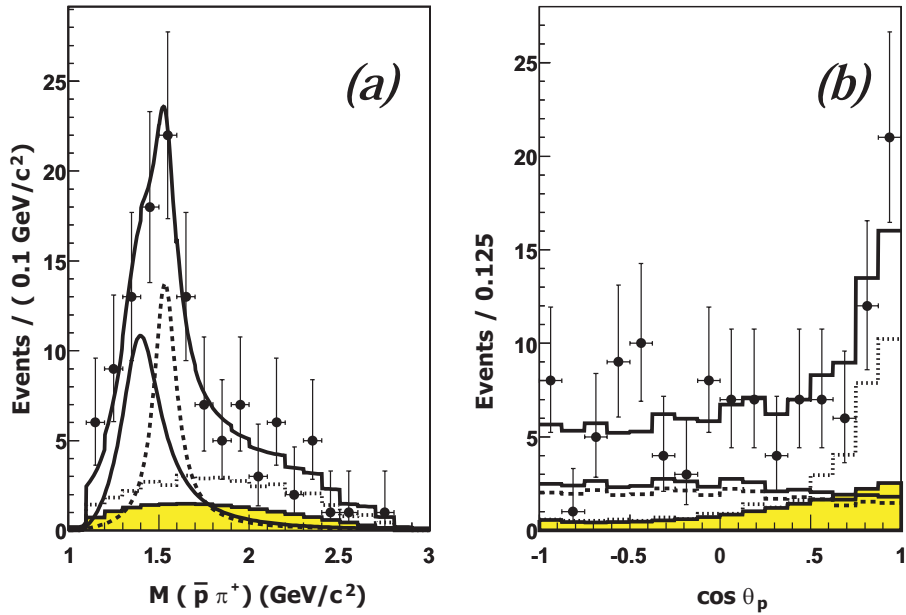}
\centering
\caption{
Simultaneous fit to the $M(\bar{p}\pi^+)$ and $\cos{\theta}_{\bar p}$ distributions 
with the $N(1440)P_{11}$ and $N(1535)S_{11}$ Breit-Wigners. 
The histograms indicate the contributions from the $P_{11}$ (solid) and $S_{11}$ (dashed) states,
{\MCC} (dotted), and the background (shaded). 
}
\label{bwfit2}
\end{figure}

Finally, we try to perform a fit with an incoherent sum of the two Breit-Wigners,
as we find that the fit results favor $N(1440)P_{11}$ and $N(1535)S_{11}$,
and both give a distribution uniform in $\cos\theta_p$.
Figure~\ref{bwfit2} shows the result of a fit to (a) the $M(\bar{p}\pi^+)$ and (b) $\cos{\theta}_{\bar p}$ distributions,
where the $N$ masses and widths are fixed to the values in Ref.~\cite{pdg2006}, and
the individual yields are floated. The histograms show the contributions from $N(1440)$ (solid),
$N(1535)$ (dashed) states, {\MCC} (dotted), and the background (shaded).
The yields are $(37\pm12)$ for the $N(1440)$ and $(30\pm11)$ for the $N(1535)$, 
while the {\MCC} yield is $(35\pm9)$. 
We obtain the goodness of fit $\chi^2$/ndf=30.3/33, which
indicates a slight preference (by 2.7\,$\sigma$) for a mixed state 
of $N(1440)$ and $N(1535)$~\cite{pdg2006}.

In summary, we study the three-body decay
$\bar{B}^0\rightarrow\Sigma_c(2455)^{0}\bar{p}\pi^{+}$
with the same data set used for the analysis of the four-body decay
$\bar{B}^0\rightarrow\Lambda_c^{+}\bar{p}\pi^{+}\pi^{-}$~\cite{park}.
We observe a broad $\bar{p}\pi^+$ mass structure near $1.5\,{\rm GeV}/c^2$, and
a uniform $\cos{\theta}_{\bar p}$ distribution with a sharp forward peak. 
To explain these structures, we find that contributions from 
an intermediate two-body decay {\MCA}, 
non-resonant three-body decay {\MCB} and a low mass structure near threshold {\MCC} are needed. 
We perform a simultaneous fit to the $M(\bar{p}\pi^+)$ and $\cos{\theta}_{\bar p}$ distributions
with those three modes, and determine the yield and
the relativistic Breit-Wigner parameters of the ${\bar N}^0$ state for {\MCA}.
We obtain the branching fraction product of {\brproduct}
with a signal significance of {\ssgnfn} standard deviations including systematics.
The fitted mass and width are consistent with $\bar{N}(1440)$$P_{11}$ and $\bar{N}(1535) S_{11}$;
both states also produce a uniform helicity distribution that is in good agreement with the data.
The structure is also consistent with an interpretation in terms of an admixture of these two states.

\vspace {0.5cm}
ACKNOWLEDGMENT

\vspace {0.5cm}

We thank the KEKB group for excellent operation of the
accelerator, the KEK cryogenics group for efficient solenoid
operations, and the KEK computer group and
the NII for valuable computing and Super-SINET network
support.  We acknowledge support from MEXT and JSPS (Japan);
ARC and DEST (Australia); NSFC (China); 
DST (India); MOEHRD, KOSEF and KRF (Korea); 
KBN (Poland); MES and RFAAE (Russia); ARRS (Slovenia); SNSF (Switzerland); 
NSC and MOE (Taiwan); and DOE (USA).

\end{document}